\documentclass[aps,prb,
12pt,
]{revtex4-1}
\usepackage{bm}
\usepackage{amsmath}    
\usepackage{amssymb}    
\usepackage{graphicx}   
\usepackage{booktabs}
\usepackage{xcolor}
\usepackage[hang]{subfigure}
\usepackage{hyperref}
\DeclareMathOperator{\sech}{sech}

\newcommand{\jun}{junction }
\newcommand{\juns}{junctions }
\newcommand{\Jos}{Josephson }

\newcommand{\conf}{confocal }

\begin{document}
\title[Roberto Monaco]{Flux Flow Effects in Annular Josephson Tunnel Junctions}
\author{Roberto Monaco}
\email[Corresponding author e-mail address:]{  r.monaco@isasi.cnr.it and roberto.monaco@cnr.it}
\affiliation{CNR-ISASI, Institute of Applied Sciences and Intelligent Systems ''E. Caianello'', Comprensorio Olivetti, 80078 Pozzuoli, Italy and\\ International Institute for Advanced Scientific Studies (IIASS), Vietri sul Mare, Italy}

\author{Jesper Mygind}
\affiliation{DTU Physics, B309, Technical University of Denmark, DK-2800 Lyngby, Denmark}
\email{myg@fysik.dtu.dk}

\author{Valery P. Koshelets}
\affiliation{Kotel'nikov Institute of Radio Engineering and Electronics,
Russian Academy of Science, Mokhovaya 11, Bldg 7, 125009 Moscow, Russia.}
\email{valery@hitech.cplire.ru}

\date{\today}

\begin{abstract}
\vskip -24pt
We investigate Josephson flux-flow in annular Josephson tunnel junctions (AJTJs) under the application of magnetic fields generating finite-voltage steps in their current-voltage characteristics. Experimental data are presented for confocal AJTJs which are the natural generalization of the well studied circular AJTJs for which flux flow effects have never been reported. Displaced linear slopes, Fiske step staircases and Eck steps were sequentially recorded at $4.2\,K$ with high-quality $Nb/Al$-$AlOx/Nb$ confocal AJTJs when increasing the strength of a uniform magnetic field applied in the plane of the junction. Their amplitude was found to strongly depend not only on the strength, but also on the orientation, of the external field. Extensive numerical simulations based on a phenomenological sine-Gordon model developed for confocal AJTJs were carried out to disclose the basic flux-flow mechanism responsible for the appearance of magnetically induced steps and to elucidate the role of several critical parameters, namely, the field orientation, the system loss and the annulus eccentricity. It was found that in a topologically closed system, such as the AJTJ, where the number of trapped fluxons is conserved and new fluxons can be created only in the form of fluxon-antifluxon pairs, the existence of a steady viscous flow of Josephson vortices only relies on the capability of the fluxons and antifluxons to be generated and to annihilate each other inside the junction. This also implies that flux-flow effects are not observable in circular AJTJs.


\end{abstract}
\maketitle
\tableofcontents
\newpage

\section{Introduction}

During the last decades, the unidirectional motion of a train of Josephson supercurrent vortices, called \textit{Josephson flux-flow} (JFF), has attracted intensive theoretical and experimental interest \cite{Barone71,Rajeevakumar80,Nagatsuma83,Zhang93,Koshelets97,Bulaevskii06,Gulevich17}. The investigations of the JFF were focused on rectangular planar Josephson Tunnel Junctions (JTJs) in the presence of a static in-plane magnetic field. The most studied geometrical configuration is the one-dimensional junction with one dimension longer and the other much shorter than the Josephson penetration length, $\lambda_J$. The external magnetic field applied in the junction's plane and perpendicular to the long dimension penetrates from both extremities of the long JTJ and creates a distributed static chain of \Jos vortices, so called \textit{fluxons} as each of them carries one magnetic flux quantum. The fluxon density along the chain increases with the field strength. When the junction is biased with a dc current the chain starts to move until it reaches a steady velocity that increases with the bias but never exceeds the Swihart velocity \cite{Swihart}, $\bar{c}$, which is the characteristic speed of electromagnetic waves in JTJs (typically a few percent of the free-space velocity). In this regime fluxons are created at one boundary of the junction and annihilate at the other boundary where they emit electromagnetic radiation. The unidirectional and viscous flow of magnetic flux quanta in a long overlap-type JTJ has been successfully employed to realize tunable sub-millimeter-wave oscillators, called \textit{flux-flow oscillators} (FFOs), whose radiation frequency is determined by the spacing between the moving fluxons and the velocity of the fluxon chain. In the past few years the FFOs were developed to the stage of practical applications both on board of high-altitude balloons and in the laboratory \cite{Lange10,Li12,Sobak17,Gulevich19}. The fluxon train is not a rigid array of vortices and has its internal degrees of freedoms; as local variations of the fluxon spacing change the radiation frequency, the flux-flow steadiness determines the line-width of the emitted CW radiation. The back-reflected radiation (so-called \textit{plasma wave}) may interact with the incident fluxons, in particular at low chain densities \cite{Nagatsuma85}. In order to optimize the output power and to minimize the unwanted backward radiation, special geometrical configurations have been implemented that improve the impedance matching to the RF circuit connected to the FFO at the junction end where the fluxons annihilate and radiation is emitted \cite{Nagatsuma985}. In addition, also the loss in the junction is a critical parameter in the JFF stability; large dissipation damps both the plasma waves and the flow of magnetic energy and, vice-versa, in an underdamped system the fluxon motion becomes irregular especially when the fluxon speed approaches the limiting velocity, $\bar{c}$. 


\medskip

\begin{figure}[t]
\centering
\includegraphics[width=8cm]{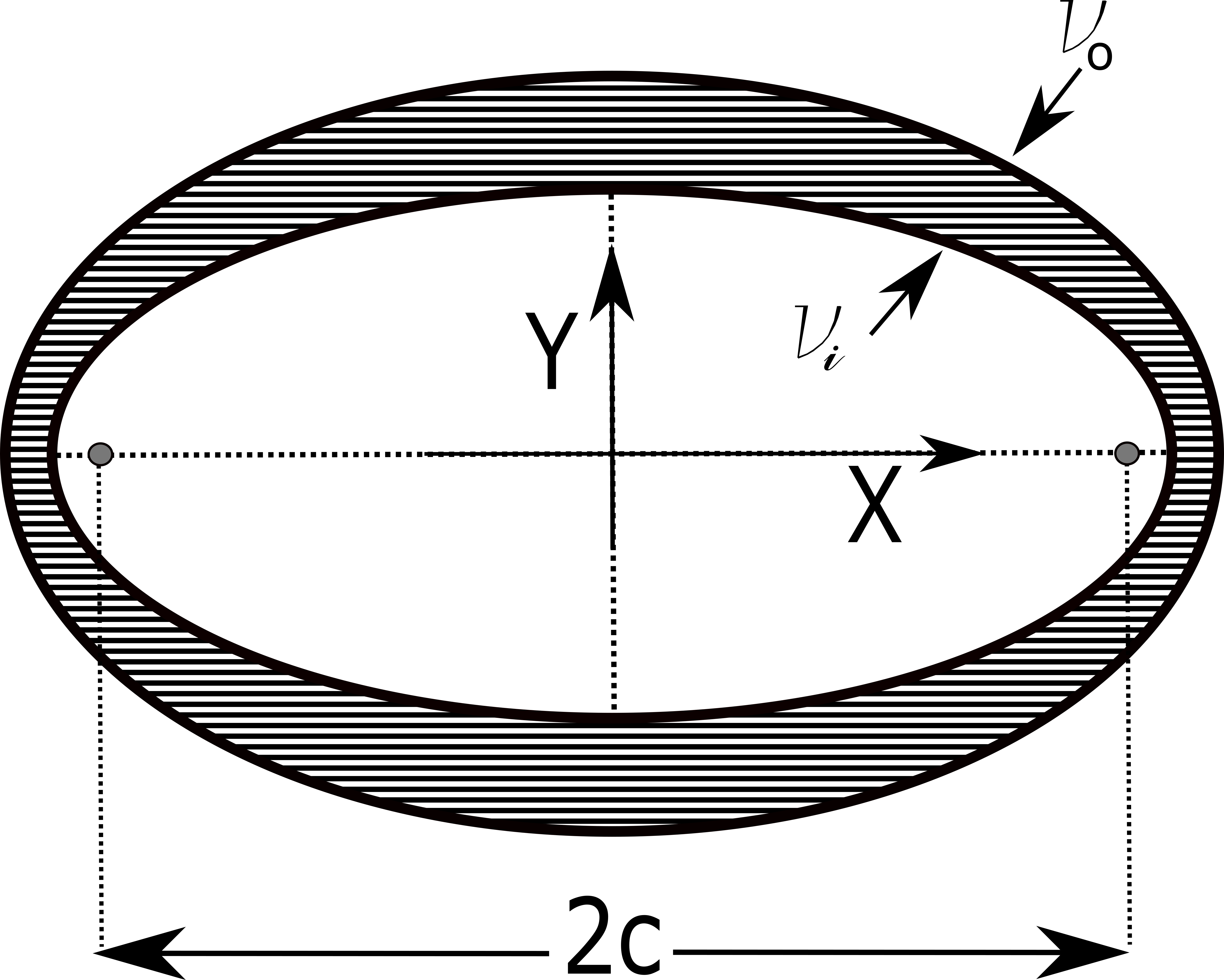}
\caption{Drawing of a \textit{confocal} annulus delimited by two closely spaced ellipses having the same foci - the gray dots. The hatched area represents the tunneling area of a confocal annular Josephson tunnel junction. The inner and outer elliptical boundaries are uniquely determined by their radial elliptic coordinate, respectively, $\nu_i$ and $\nu_o<\nu_i$. As the foci move towards the origin, the eccentricity vanishes and the confocal annulus progressively reduces to a circular annulus (with uniform width).}
\label{ConfAnn}
\end{figure}

\noindent Beside the rectangular simply-connected geometry, another one-dimensional configuration has been successfully used to study the fluxons propagation and to experimentally test the perturbative sine-Gordon models developed to take into account the dissipative effects. It is the annular geometry in which the JTJ is a the superposition of two narrow doubly-connected superconducting electrodes. In this configuration the influence of the end-boundaries is avoided, as the boundary conditions of the open simply-connected configuration are replaced by periodic conditions. Due to the fluxoid quantization in a superconducting loop \cite{mercereau63}, the numerical imbalance of the magnetic flux quanta trapped in the doubly-connected electrodes of the annular JTJ (AJTJ) during the normal-superconducting transition leads to a different winding number of vorticity and thus the appearance of a number of \Jos vortices trapped in the tunnel barrier. The AJTJ is a topologically closed system such that the number of trapped fluxons is conserved and new fluxons can be created only in the form of fluxon-antifluxon ($F\bar{F}$) pairs. When the ring-shaped electrodes are narrower than $\lambda_J$ and their curvature is everywhere much larger than $\lambda_J$, then the motion of a single fluxon along the perimeter of an AJTJ can be assimilated to that on an infinite structure. The simplest and most studied annular geometry has been implemented with \textit{circular} AJTJs realized by the superposition of two concentric circular annuli \cite{davidson85, dueholm,hue}; in this configuration a magnetic field applied in the junction plane gives rise to a tunable sinusoidal periodic potential for the trapped fluxon \cite{gronbech, ustinov,PRB98,JLTP01,wallraff03}. It has been shown that the fluxon energy levels are quantized when cooled to milli-Kelvin temperatures \cite{wallraff00}. Despite the many theoretical and experimental investigations on circular AJTJs, the phenomenon of flux-flow has never been reported which implicitly suggests that a regular motion of a fluxon chain is impeded by the periodic boundary conditions. 

\noindent Recently, the circular geometry in which the internal and external boundaries of the annulus are closely spaced concentric circumferences has been generalized to the so-called \textit{\conf geometry} in which the annulus boundaries are confocal ellipses \cite{JLTP16b,JPCM16}, rather than concentric circles. The circular AJTJs can be seen as a special case of the \textit{\conf} AJTJ (CAJTJ) where the elliptic boundaries have zero eccentricity. Since the physics of Josephson planar tunnel junctions drastically depend on their geometrical configurations \cite{Barone} and even tiny geometrical details can play a determinant role \cite{JLTP17}; it is not surprising that the CAJTJs have a very rich nonlinear phenomenology that strongly depends on the system eccentricity \cite{SUST18,JLTP18}. The key ingredient of this geometrical configuration is the periodically varying barrier width that generates an intrinsic spatially dependent potential for the vortex with bistable states. The two-state vortex potential can be fine-tuned by an in-plane magnetic and a reliable manipulation of the vortex state. This key ingredient for the realization of a quantum bit has been classically demonstrated in CAJTJs \cite{SUST18,JLTP18}. In addition, the confocal annular configuration is very well modeled by a modified and perturbed one-dimensional sine-Gordon equation that admits solitonic solutions. The tunneling area of a CAJTJ is drawn in Fig.~\ref{ConfAnn} where the principal diameters of the closely spaced ellipses with the same interfocal separation, $2c$, are parallel to the $X$ and $Y$ axes of a Cartesian coordinate system. The common foci, the gray dots at $(\pm c,0)$, lie on the $X$-axis. In elliptical coordinates all possible confocal ellipses are uniquely identified by a characteristic value $\nu_c>0$. If we name $\nu_i$ and $\nu_o<\nu_i$ the radial parameters of, respectively, the inner and outer ellipses, then $\Delta \nu\equiv\nu_o-\nu_i$ measures the separation between the ellipses. The annulus is narrow if $\Delta \nu<\bar{\nu}\equiv(\nu_o+\nu_i) /2$. For such an annulus the mean value, $\bar{\nu}$, is related to its aspect ratio, $\rho$, defined as the ratio of the mean length of the minor axes to the mean length of the major axes, $\rho\equiv\tanh\bar{\nu}\leq 1$, and to its eccentricity, $e^2 \equiv 1-\rho^2=\sech^2\bar{\nu}\leq 1$. It is worth to stress that two ellipses can never be ''parallel'', therefore, in general, a confocal annulus has an intrinsic non-uniform width. The width of the confocal annulus is smallest at the equatorial point, $\Delta w_{min}$, and largest at the poles, $\Delta w_{max}$; the width variation is smoothly distributed along one fourth of the perimeter, $L$, of the confocal annulus. In the limiting case of a vanishing eccentricity, the foci of the ellipse collapse to a point at the origin (i.e., $c\to0$) and the ellipse turns into a circle. At the same time, $\cosh\nu_c$ diverges, while the product $c \cosh\nu_c$ remains finite and tends to the radius, $r$, of the circle. A circular annulus has unitary aspect ratio, zero eccentricity and uniform width. The confocal AJTJs should not be confused with the \textit{elliptical} AJTJs \cite{Peterson,SUST15,nappimonaco,JLTP16a} whose internal and external boundaries are closed curves parallel to a master ellipse, with opposite offsets; strictly speaking, such curves are not ellipses, but more complex curves. 

\noindent Generally speaking, the motion of Josephson vortices along a current biased JTJ is manifested by stable current branches or singularities in its current-voltage characteristic (IVC) at a finite voltage proportional to the fluxon number and their time-averaged speed. In the absence of an external magnetic field these current singularities are called the Zero-Field Steps (ZFSs) and correspond to the motion of just one or a few particle-like flux quanta along the extended dimension of the junction. In the presence of a magnetic field either externally applied or self-induced by the bias current, several families of singularities can appear on the junction IVC, Displaced Linear Slopes (DLSs), Fiske Steps (FSs) or Eck Steps (ESs), corresponding to different dynamical states \cite{Cirillo98}. In this paper we report on an extensive experimental investigation of the IVCs carried out on high-quality $Nb/Al$-$AlOx/Nb$ AJTJs under a large variety of conditions; it was found that the JFF is possible in long AJTJs and its effects are more pronounced when the annulus is confocal and has a large eccentricity. Our findings are supported by systematic numerical simulations that provide the details of the JFF dynamic properties in long AJTJs. 

\subsection{Outline of the paper}
\vskip 5pt
\noindent The paper is organized into four sections. Sec.II contains the experimental findings: we first describe the electrical and geometrical features of our low-loss $Nb/Al$-$AlOx/Nb$ window-type long CAJTJs all having the same circumference, but different eccentricity; later on we present and comment on their IVCs recorded at $4.2\,K$ for different values of an externally applied in-plane magnetic field. In Sec.III we introduce the theoretical framework for the study of a current-biased CAJTJ subjected to an external magnetic field in the framework of a modified and perturbed sine-Gordon equation; we then present numerically calculated IVCs with parameters taken from the experiments and describe the dynamical state in the flux-flow regime. The numerical results are compared with experiment, and good agreement is found in most cases. Some comments and the conclusions of our work are presented in Sec. IV.

\section{The measurements}

\subsection{The samples and the experimental setup}

\begin{figure}[b]
\centering
\includegraphics[height=6.5cm]{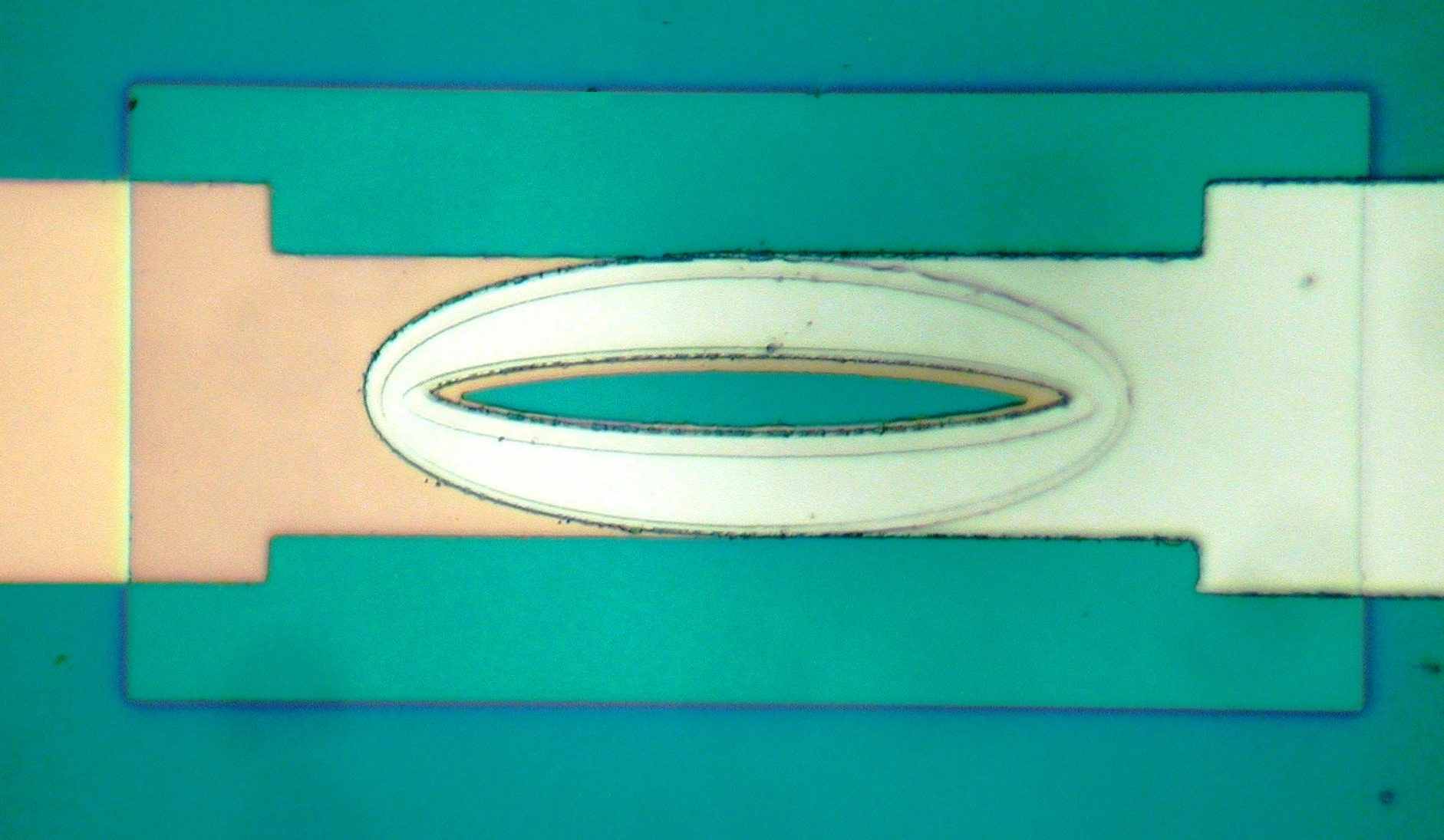}
\caption{(Color online) Optical image of a \textit{Lyngby-type} confocal annular Josephson tunnel junction (CAJTJ) made by the superposition of two $Nb$ doubly-connected electrodes. For this sample the ratio of the minor axis and the major axis is $1\!:\!4$ that implies that the equatorial annulus width is one forth of the polar width. The DC bias current flows in the two horizontal electrodes.}
\label{picture}
\end{figure}

\noindent The CAJTJs used for our investigation were fabricated using the well known and reliable selective niobium etching and anodization process\cite{sneap}. A $30\, nm$ thick Al oxide obtained from liquid anodization \cite{JAP95} and an extra $230\, nm$ thick dielectric layer made of rf-sputtered silicon dioxide provide the electrical insulation between the base electrode and the wiring film around the \jun area. The details of the $Nb/Al$-$AlOx/Nb$ trilayer deposition and of the fabrication process can be found elsewhere \cite{VPK,Filippenko}. Two batches were made using different oxidation times of the $Al$ overlayer yielding samples with quite different critical current densities, $J_c$. All our samples were designed with the so-called \textit{Lyngby-type} geometry\cite{davidson85} that refers to a specularly symmetric configuration in which the width of the current carrying electrodes matches one of the ellipse outer axis. One example of this geometry is shown in Fig.~\ref{picture}. The bias current flows parallel to the major axis of the confocal annulus. A Molybdenum resistive film was integrated on the chip for fast and reliable heating of the samples. More details of the chip layout and design were reported in Ref.\cite{SUST18}.

\medskip

\noindent Our setup consisted of a cryoprobe inserted vertically in a commercial $LHe$ dewar. The $Si$ chip with the CAJTJs is mounted on a Cu block enclosed in a vacuum-tight can immersed in the liquid He bath. The cryoprobe was magnetically shielded by means of two concentric superconducting $Pb$ cans surrounded by a long cryoperm can; in addition, the measurements were carried out in an rf-shielded room. The chip was positioned in the center of a long superconducting cylindrical solenoid whose axis was along the vertical direction to provide an in-plane magnetic field, either parallel, $H_{\parallel}$, or perpendicular, $H_{\bot}$, to the annulus major axis depending on the junction orientation. All the experiments reported in this work were carried out in the flux free regime, i.e., with no fluxon trapped in the AJTJs at the time of their normal-to-superconducting transition.


\medskip

\noindent A large number of CAJTJs were investigated having different geometrical and electrical parameters but the same mean circumference $L= 4c \cosh\bar{\nu}\,\texttt{E}(e^2)=200 \mu m$, where $\texttt{E}(e^2)\equiv \texttt{E}(\pi/2,e^2)$ is the {\it complete} elliptic integral of the second kind of argument $e^2$. All the samples showed highly hysteretic IVCs with low subgap leakage currents, $I_{sg}$, compared to the current jump at the gap voltage, $\Delta I_g$; this means that the \juns have high quality and are strongly underdamped. In addition, they all showed a maximum critical current, $I_c^{max}$, being considerably smaller than about the $\approx 70\%$ of $\Delta I_g$, typical of short $Nb/Al$-$AlOx$-$Al/Nb$ junctions. This indicates a non-uniform bias current distribution and, more importantly, the presence of so-called \textit{self-field effects} \cite{SUST13a,SUST15}. Nominally identical samples made within the same fabrication run gave qualitatively similar results; the findings presented in this work pertain to just two representative ones having $J_c\approx 4.7 kA/cm^2$ that corresponds to $\lambda_J\approx 4.0 \mu m$. The geometrical details of the tunneling area for the selected CAJTJs and their relevant electrical parameters (measured at $4.2\,K$) are listed in Table I. Their DC current-biasing electrodes were parallel to the annulus major diameter, as shown in Fig.~\ref{picture}. Essentially the two samples in Table I differ by their aspect ratio, $\rho=1/4$ and $1/2$. They were designed to have the same equatorial width, $\Delta w_{min}=2.1 \mu m$, so that the annulus polar width, $\Delta w_{max}=\Delta w_{min}/\rho$, is twice as large in the first sample that, consequently, also has a large area. More specifically, the areas, $\pi c^2 \cosh2\bar{\nu}\Delta\nu$, of the two CAJTJs happen to be approximately in the ratio $2:1$, similar to the ratio of their current jumps at the gap voltage, $\Delta I_g$. Both our specimens have the same normalized perimeter, $\ell\equiv L/\lambda_J\approx50$. We like to stress that the tunneling area of a CAJTJ, regardless of the geometry of the current carrying electrodes, is uniquely determined once the interfocal distance, $2c$, the aspect ratio, $\rho$, and $\Delta \nu$ are given.

\begin{table*}[t]
	\centering
	 	\begin{tabular}{|c||c|c|c|c|c|c|c|c|c|c|}
		 \hline
     $\rho$ & $\bar{\nu}$ &  $\Delta \nu$ & $c$ & $\Delta w_{min}$ & $\Delta w_{max}$ & $Area$ &$\Delta I_g$ & $I_{sg}(2mV)$& $I_c^{max}$ & $H_{\bot}^c$ \\
		\hline
		 &  & & $\mu m$ & $\mu m$ & $\mu m$ & $\mu m^2$ & $mA$ & $mA$ & $mA$ & $mT$\\
		\hline
    	1/4 & 0.26 & 0.18  & 45.1 & 2.1 & 8.4  & 1310 &94 & 4.8 & 26 & 0.91\\
			\hline
    	 1/2 & 0.55 & 0.10  & 35.8 & 2.1 & 4.2  & 680 &48 & 2.6 & 21 & 0.90\\
		 \hline
				\end{tabular}
		\caption{Some geometrical details and electrical parameters (measured at $4.2\,K$) of two representative CAJTJs with the same critical current density, $J_c\approx4.7\,kA/cm^2$ (corresponding to $\lambda_J\approx4.0 \mu m$), and the same mean perimeter, $L=200 \mu m$, but different values of the aspect ratio, $\rho$, (either $1/4$ or $1/2$). $\Delta \nu=\Delta w_{min}/c \sinh \bar{\nu}$.}
		 \label{table0}
\end{table*}

\medskip


\subsection{Current-voltage characteristics}

We now present the evolution of the current-voltage characteristics obtained by sweeping the bias current with a triangular waveform on our CAJTJs subject to a uniform in-plane magnetic field . At zero and very small magnetic-field strength, fluxon-antifluxon pairs are nucleated in the low voltage region of the IVC and \textit{zero-field steps} (ZFSs) are observed. Their position depends on the number of nucleated pairs and on the mean propagation velocity of the fluxons and antifluxons along the junction perimeter. In this dynamical state, driven by the Lorentz force generated by the bias current, the fluxons and the antifluxons travel in opposite directions in an intrinsic spatially periodic potential due to the variable width of the CAJTJ. The unidirectional motion of a single vortex in CAJTJs and its resonant interaction with the plasma waves have been studied both experimentally and numerically in Ref.\cite{SUST18}.

\begin{figure}[b]
\centering
\subfigure[ ]{\includegraphics[height=6cm,width=8cm]{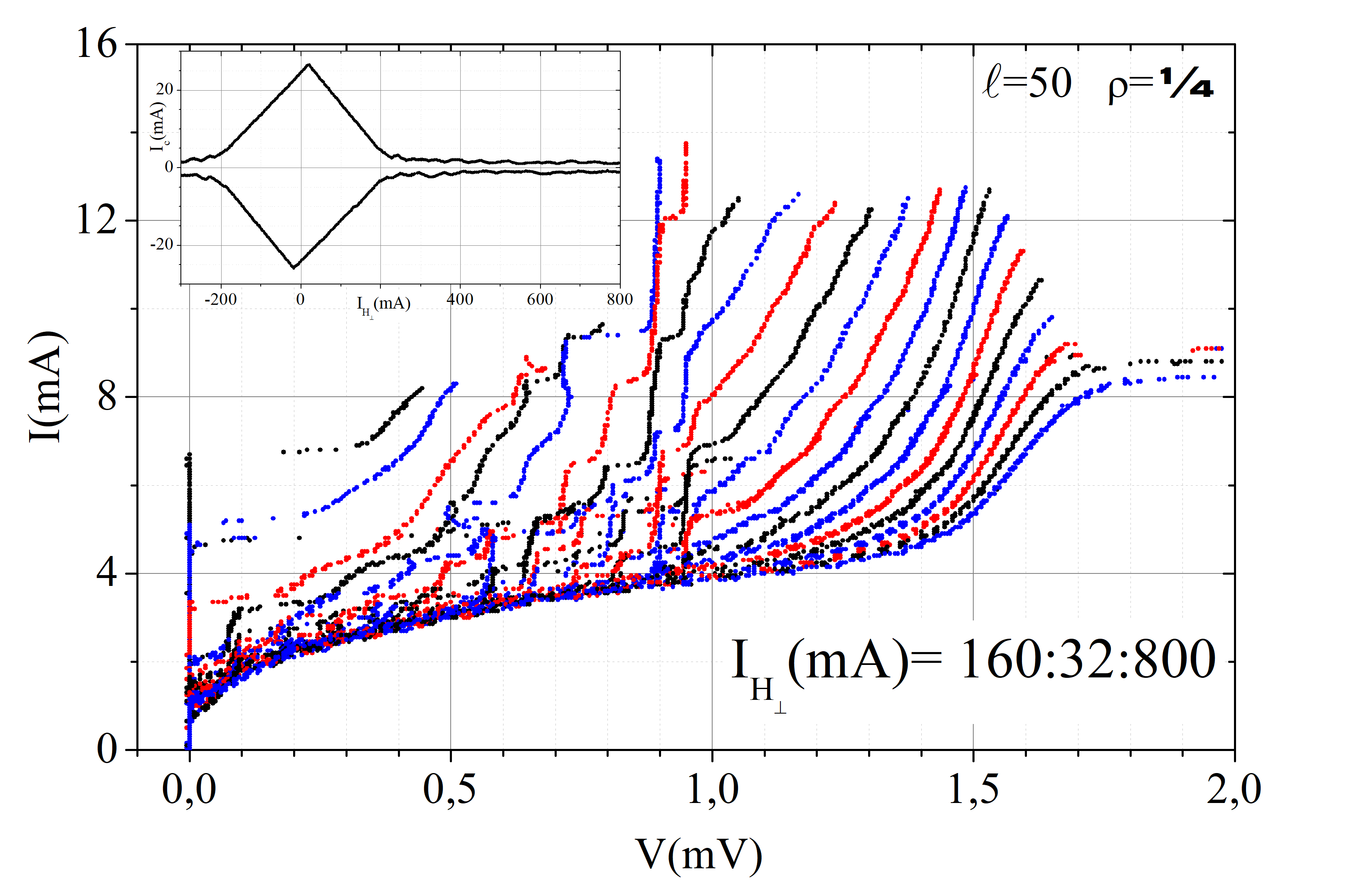}}
\subfigure[ ]{\includegraphics[height=6cm,width=8cm]{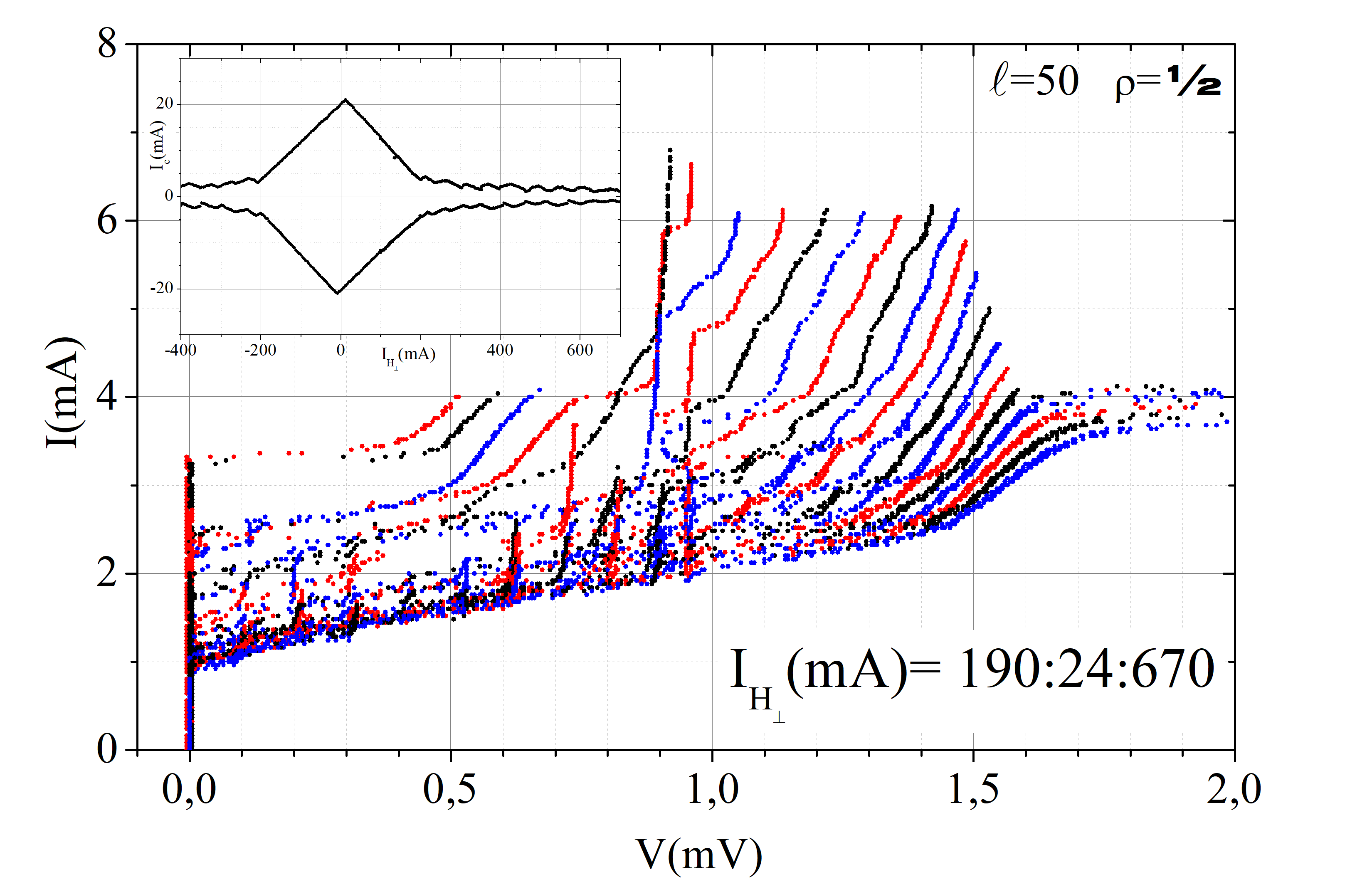}}
\caption{(Color online) I-V characteristics of the two CAJTJs listed in Table I recorded at different values of an in-plane magnetic field, $H_{\bot}$, produced by a control current, $I_{H_{\bot}}$, in a superconducting solenoid in the direction perpendicular to the major axis of the confocal annuli. The CAJTJs have the same normalized perimeter, $\ell=L/\lambda_J=50$, but different aspect ratios, $\rho$: a) $\rho=1/4$ with the solenoid current in the range $160$\textendash $800\, mA$ with the increment of $32\,mA$; b) $\rho=1/2$ with the solenoid current in the range $190$\textendash$\,670\, mA$ with the increment of $24\,mA$. The solenoid field-to-current conversion factor is $3.9\, \mu$T/mA. The insets show the corresponding magnetic diffraction patterns of the zero-voltage critical current, $I_c(H_{\bot})$.}
\label{IVC}
\end{figure}

\noindent As far as concerns the occurrence of current singularities induced by an externally applied in-plane magnetic field, two critical parameters were recognized: i) the orientation of the magnetic field and ii) the sample aspect ratio. As the ellipse has two axes of symmetry, it is expected that the response of a CAJTJ to the in-plane magnetic field is strongest when the magnetic field is perpendicular to the major axes, as it occurs in elliptical JTJs \cite{Peterson}. It has been reported that for CAJTJS the magnetic diffraction patterns (MDPs) of the zero-voltage critical current, $I_c(H)$, obtained with a field perpendicular, $H_{\bot}$, and parallel, $H_{\parallel}$ to the major axis, differ from one another not only quantitatively but also qualitatively \cite{SUST15}. The perpendicular MDP, $I_c(H_{\bot})$, shows a fast initial suppression of the critical current and small secondary lobes; vice-versa, $I_c(H_{\parallel})$ is characterized by a slow modulation of the critical current and by large secondary lobes. Interestingly, no current singularities were recorded on the IVCs of a CAJTJ subjected to a even large parallel field; on the contrary, a large variety of current branches appeared using a perpendicular field. The families of IVCs of the two CAJTJs in Table I recorded at different values of the perpendicular magnetic fields produced by the control current, $I_{H_{\bot}}$, in the superconducting solenoid are presented in Figs.~\ref{IVC}(a) and (b). The field-to-current conversion factor of the solenoid close-fitting in the superconducting shields \cite{SUST09} is $3.9\, \mu$T/mA. The insets show the corresponding MDPs with the horizontal scale expressed in terms of the control current $I_{H_{\bot}}$. The moderate skewness of the MDPs is ascribed to the fact that for both samples the bias current flow occurs in the direction orthogonal to the applied field \cite{SUST13a}; in this configuration the magnetic field induced by the measuring current (self-field) adds to the external field in the second and fourth quadrants, while in the first and third quadrants it partially compensates the applied field. The two samples happen to have almost the same perpendicular critical field, $H_{\bot}^c$, as reported in the last entry of Table I. 

\noindent The two families of IVCs in Figs.~\ref{IVC}(a) and (b) have the same qualitatively features. At magnetic fields smaller that the (perpendicular) first critical field, $H_{\bot}^c$, the so-called \textit{displaced linear slope} (DLS) appears, first observed in large-$I_c$ square JTJs having cross type geometry \cite{Scott69} and soon recognized to be a manifestation of flux flow \cite{Barone71}. Upon increasing $H_{\bot}$, the DLS branch shifts almost linearly with the field strength towards higher voltages \cite{Rajeevakumar80}. In our samples the DLS is not quite linear, however, its differential resistance is almost constant as a function of the current and magnetic field. As $H_{\bot}$ approaches $H_{\bot}^c$, seamlessly an additional, more vertical, branch develops on the top of the DLS made by the hysteretic superposition of a series of quantized steps, the so-called \textit{Fiske steps} (FSs), originating from the cavity resonant interaction between the alternating Josephson current and the electromagnetic fields \cite{Fiske65}. When $H_{\bot}$ exceeds $H_{\bot}^c$ we shall call these steps \textit{flux-flow steps} (FFSs). Their asymptotic voltage increases with the field strength and their splitting in sub-steps takes place up to a specific ‘‘boundary’’ voltage, $V_b\approx 900 \mu V$, where the FFS switching current is largest and its differential resistance is smallest. As seen in Figs.~\ref{IVC}(a) and (b), for $V>V_b$, all the FSs merge in a single smooth singularity \cite{Cirillo98}, also called an \textit{Eck step} \cite{Eck}, whose voltage, for a fixed current, increases linearly upon the value of the external magnetic field up to the gap voltage. The ‘‘boundary’’ voltage has been observed in linear JTJs with high current density ($J_c>1 kA/cm^2$) and has been explained by the effect of Josephson self-coupling which is due to the absorption of AC Josephson radiation energy by the quasi-particles \cite{Koshelets97,Gulevich17,Koshelets19}. From a quantitative perspective, the comparison of the two families of IVCs in Figs.~\ref{IVC}(a) and (b) shows that, the amplitudes of the FFSs, once the background quasiparticle current is subtracted, almost scale with the junctions areas.

\medskip

\noindent Families of IVCs qualitatively similar to those in Figs.~\ref{IVC}(a) and (b) were recorded in samples with the same geometrical configuration but with a lower critical current density, $J_c\approx 2.2 kA/cm^2$, that corresponds to a larger \Jos penetration depth \cite{SUST18}, $\lambda_J\approx 6.2 \mu m$, and, therefore, to a smaller normalized length, $\ell \approx 32$. Apart from an obvious scale factor, the main relevant difference was identified as the dependence of the step heights on the junction aspect ratio. More specifically, for our low-$J_c$ CAJTJs, the magnetically induced branches were less pronounced in samples with smaller eccentricities.

\section{Theory of one-dimensional confocal AJTJs} 

\subsection{The sine-Gordon model} 


For many decades the sine-Gordon model has been the most adequate model for the JTJ, giving a good qualitative description of its basic properties, such as Fiske steps, vortices dynamics, etc. In this phenomenological model the electrodynamics of a long JTJ in the presence of magnetic field and losses is described by the perturbed sine-Gordon equation \cite{Barone}. The geometry of our system suggests the use of the (planar) elliptic coordinate system $(\nu,\tau)$, a two-dimensional orthogonal coordinate system in which the coordinate lines are confocal ellipses and hyperbolae. In this system, any point $(x,y)$ in the $X$-$Y$ plane is uniquely expressed as $(c\cosh\nu\sin\tau, c\sinh\nu\cos\tau)$ with $\nu\geq0$ and $\tau\in[-\pi,\pi]$ for a given positive $c$ value. According to these notations, the origin of $\tau$ lies on the positive $Y$-axis and increases for a clockwise rotation (Refer to Fig.~\ref{ConfAnn}). In the limit $c\to0$, the elliptic coordinates $(\nu,\tau)$ reduce to polar coordinates $(r,\theta)$, where $\theta$ is the angle relative to the $Y$-axis; the correspondence is given by $\tau\to \theta$ and $c\cosh\nu\to r$ (note that $\nu$ itself will becomes infinite as $c\to0$). For closely spaced inner and outer ellipses, $\Delta\nu\equiv \nu_o-\nu_i<<1$, the expression of the local annulus width is \cite{JLTP16b}:

\vskip -8pt
\begin{equation}
\Delta w(\tau)=c\mathcal{Q}_{\bar{\nu}}(\tau)\,\Delta\nu,
\label{width}
\end{equation}

\begin{figure}[t]
\centering
\includegraphics[width=8cm]{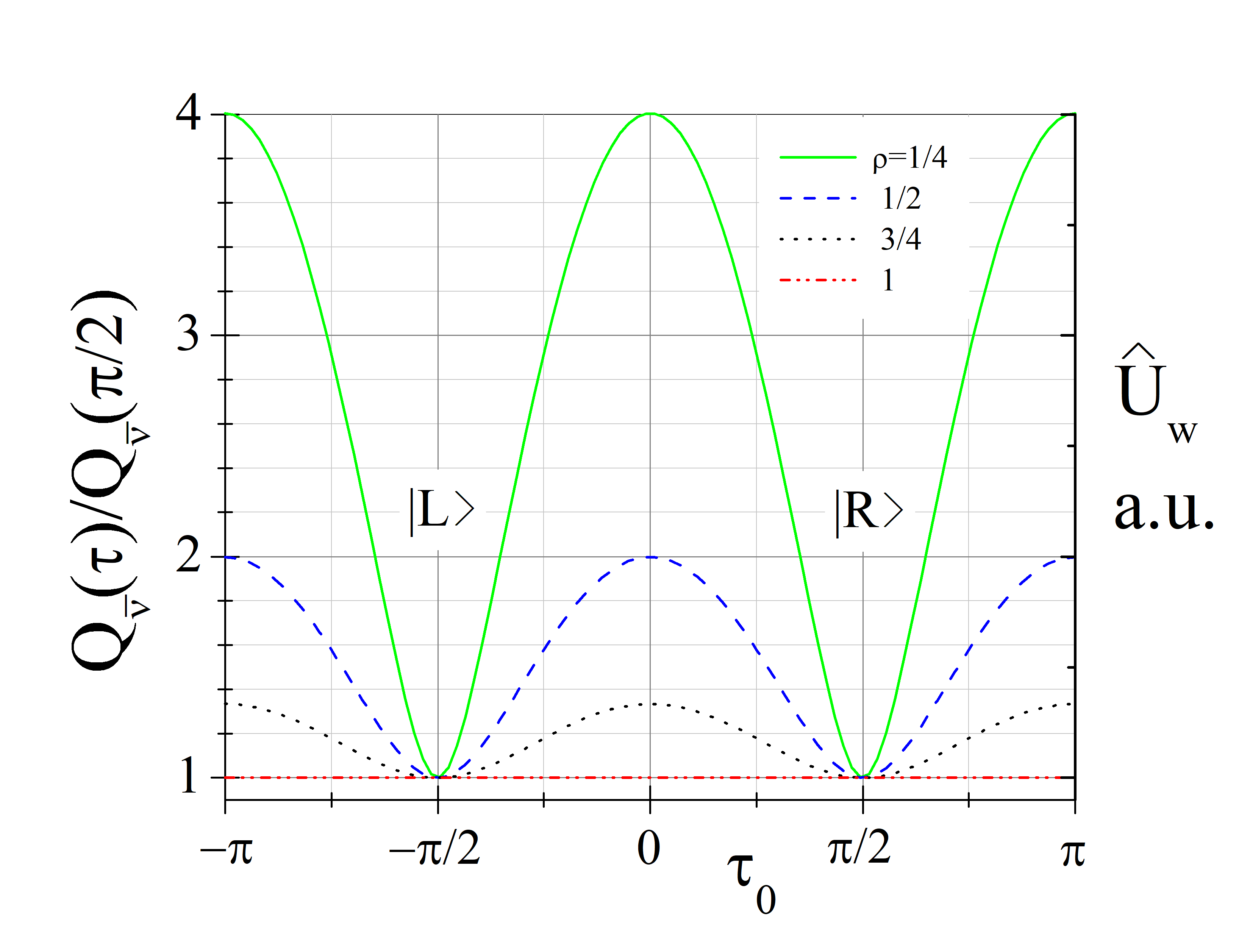}
\caption{(Color online) Dependence on $\tau$ of the normalized elliptic scale factor, $\mathcal{Q}_{\bar{\nu}}$ for several values of the aspect ratio, $\rho$. The vortex intrinsic spatially-periodic potential, $\hat{U}_w$, discussed in subsection IIIB, is proportional to $\mathcal{Q}_{\bar{\nu}}$; the left $|L\rangle$ and right $|R\rangle$ wells of the potential constitute stable classical states for the vortex with degenerate ground state energy.}
\label{Qmat}
\end{figure}

\noindent where $\mathcal{Q}_{\bar{\nu}}(\tau)$ is the elliptic scale factor defined by $\mathcal{Q}_{\bar{\nu}}^2(\tau) \equiv \sinh^2\bar{\nu} \sin^2\tau+\cosh^2 \bar{\nu} \cos^2 \tau= \sinh^2\bar{\nu}+ \cos^2\tau=\cosh^2\bar{\nu} - \sin^2\tau=(\cosh2\bar{\nu} + \cos2\tau)/2$ plotted in Fig.~\ref{Qmat} for several values of the aspect ratio, $\rho$. It is the smooth $\pi$-periodic change of the annulus width (through $\mathcal{Q}_{\bar{\nu}}$) that makes the physics of CAJTJs very rich and interesting and the modeling very accurate. The annulus width is smallest at the equatorial points ($\tau=\pm \pi/2$) and largest at the poles ($\tau=0$ or $\pm\pi$). For a circular AJTJ with unitary aspect ratio, the width is constant.

\vskip 5pt
\noindent In the small-width approximation, $\Delta w_{max}<< \lambda_J$, the system becomes one-dimensional and the $\nu$-independent \Jos phase, $\phi(\tau,\hat{t})$, of a CAJTJ in the presence of a spatially homogeneous in-plane magnetic field ${\bf H}$ of arbitrary orientation, $\bar{\theta}$, relative to the $Y$-axis, obeys a modified and perturbed 1+1 sine-Gordon equation with a space dependent effective Josephson penetration, $\lambda_J/Q_{\bar{\nu}}(\tau)$, length inversely proportional to the local junction width \cite{JLTP16b}:
\vskip -8pt
\begin{equation}
 \left[\frac{\lambda_J}{c\,\mathcal{Q}_{\bar{\nu}}(\tau)}\right]^2 \left(1+\beta\frac{\partial}{\partial \hat{t}}\right) \phi_{\tau\tau} - \phi_{\hat{t}\hat{t}}-\sin \phi =\alpha \phi_{\hat{t}} - \gamma(\tau) + F_h(\tau),
\label{psge}
\end{equation}

\noindent where $\hat{t}$ is the time normalized to the inverse of the so-called (maximum) plasma frequency, $\omega_p$, and the critical current density, $J_c$, is assumed to be uniform. The subscripts on $\phi$ are a shorthand for derivative with respect to the corresponding variable. Furthermore, $\gamma(\tau)\equiv J_Z(\tau)/J_c$ is the local normalized density of the bias current and 
\vskip -8pt
\begin{equation}
F_h(\tau)\equiv h\Delta \frac{\cos\bar{\theta}\cosh^2\bar{\nu} \sin\tau-\sin\bar{\theta}\sinh\bar{\nu}\cosh\bar{\nu} \cos\tau }{\mathcal{Q}_{\bar{\nu}}^2(\tau)}
\label{Fh}
\end{equation}

\noindent is an additional forcing term proportional to the applied magnetic field; $h\equiv H/J_c c \cosh\bar{\nu}$ is the normalized field strength for treating long CAJTJs and $\Delta$ is a geometrical factor which has been referred to as the coupling between the external field and the flux density of the annular junction \cite{gronbech}. For a \Jos ring, with $\tau$ replaced by $\theta$ and $\bar{\nu}\to \infty$, we recover the sinusoidal magnetic force \cite{PRB97}, $F_h(\theta)= h\Delta \cos(\bar{\theta}-\theta)$ with $h\equiv H/J_c r$, where $r$ is the mean ring radius. As usual, the $\alpha$ and $\beta$ terms in Eq.(\ref{psge}) account for, respectively, the quasi-particle shunt loss and the surface losses in the superconducting electrodes. 


\medskip

\noindent When cooling an AJTL below its critical temperature zero, one or more fluxons may be trapped in the AJTJ between its doubly connected electrodes \cite{PRB08}. The algebraic sum of the flux quanta trapped in each electrode is an integer number $n_w$, called the winding number, counting the number of Josephson vortices (fluxons) trapped in the \jun barrier. To take into account the number of trapped fluxons, Eq.(\ref{psge}) is supplemented by periodic boundary conditions \cite{PRB96}:
\vskip -8pt
\begin{subequations}
\begin{eqnarray} \label{peri1}
\phi(\tau+2\pi,\hat{t})=\phi(\tau,\hat{t})+ 2\pi n_w,\\
\phi_\tau(\tau+2\pi,\hat{t})=\phi_\tau(\tau,\hat{t}).
\label{peri2}
\end{eqnarray}
\end{subequations}
\vskip -4pt

\subsection{The vortex potential}

\noindent The Lagrangian and Hamiltonian densities associated with Eq.(\ref{psge}) have been derived in Ref.\cite{JPCM16}. By assuming that the annulus is long enough so that the left and right tails of a single Josephson vortex do not interact, a non-relativistic fluxon centered at $\tau_0$ is subject to an intrinsic double-well, $\hat{U}_w(\tau_0) \approx 8 \mathcal{Q}(\tau_0)$, regardless of the its polarity. Therefore, this potential applies to a single fluxon or antifluxon ($n_w=\pm1$) as well as to both the fluxon and the antifluxon of a $F\bar{F}$ pair ($n_w=0$). Referring to Fig.~\ref{Qmat}, we see that $\hat{U}_w$ expresses a $\pi$-periodic potential energy function uniquely determined by the CAJTJ ellipticity, $e^2\equiv 1-\rho^2$. The potential wells are located at equatorial point, $\tau_0=\pm \pi/2$, where the annulus width is smallest. The left $|L\rangle$ and right $|R\rangle$ wells of the potential constitute stable classical states for the vortex with degenerate ground state energy. Considering that $\sinh\bar{\nu}\leq \mathcal{Q}(\tau) \leq \cosh\bar{\nu}$, the potential wells are separated by an energy barrier that drops exponentially with $\bar{\nu}$. If a fluxon has enough energy to escape the potential wells, it starts to travel around the annulus. However, the fluxon dynamics in a CAJTJ is very different from the constant speed motion in a uniform-width circular AJTJ. In fact, the fluxon accelerates (decelerates) when it approaches the region of smallest (largest) width and other excitations such as the so-called plasma waves are radiated. Resonances may occur between the fluxon and the plasma waves \cite{pedersen86,PRB93} whose strength drastically depends on the waves amplitudes which, in turn, are strictly related to fluxon velocity and to the system's dissipation as well as to the steepness of the potential that is determined by the annulus eccentricity. The interaction between the fluxon and the small amplitude waves destabilizes its forward advancement and prevents it from reaching relativistic speeds \cite{JLTP16b}. The dispersion relation of plasma waves in confocal AJTJs has been recently investigated in the absence of trapped fluxons \cite{WM19}. It was found that for each discrete mode $m$, that corresponds to a wavelength equal to the annulus circumference divided by $m$, two eigenfrequencies exist that are related to the even and odd spatial dependence of the wave. As a result of this frequency split, the traveling wave is given by the superposition of two standing waves with the same wavelengths but different oscillation periods. Therefore, the wave profile and the velocity of the wave front are not permanent, but undergo periodic changes. 

\subsection{Numerical simulations}

The commercial finite element simulation package COMSOL MULTIPHYSICS (www.comsol.com) was used to numerically solve Eq.(\ref{psge}) subjected to the cyclic boundary conditions in Eqs.(\ref{peri1}) and (\ref{peri2}). In order to compare the numerical results with the experimental findings presented in the previous section, we set the annulus normalized length, $\ell=L/\lambda_J=50$ and the winding number, $n_w=0$, in the periodic boundary condition equal to zero (flux-free regime). We have assumed a uniform current distribution, i.e., $\gamma(\tau)= \gamma_0$. In addition, the field coupling constant, $\Delta$, was set equal to $1$. The damping coefficient $\alpha$ was changed in the weakly underdamped region $0.1 \leq\alpha\leq 0.3$, while the surface losses were simply neglected ($\beta=0$) to save computer time. CAJTJs with different values of the aspect ratio, $\rho$, where simulated to investigate the effects of the annulus eccentricity on the JFF.

\begin{figure}[t]
\centering
\subfigure[ ]{\includegraphics[width=8cm]{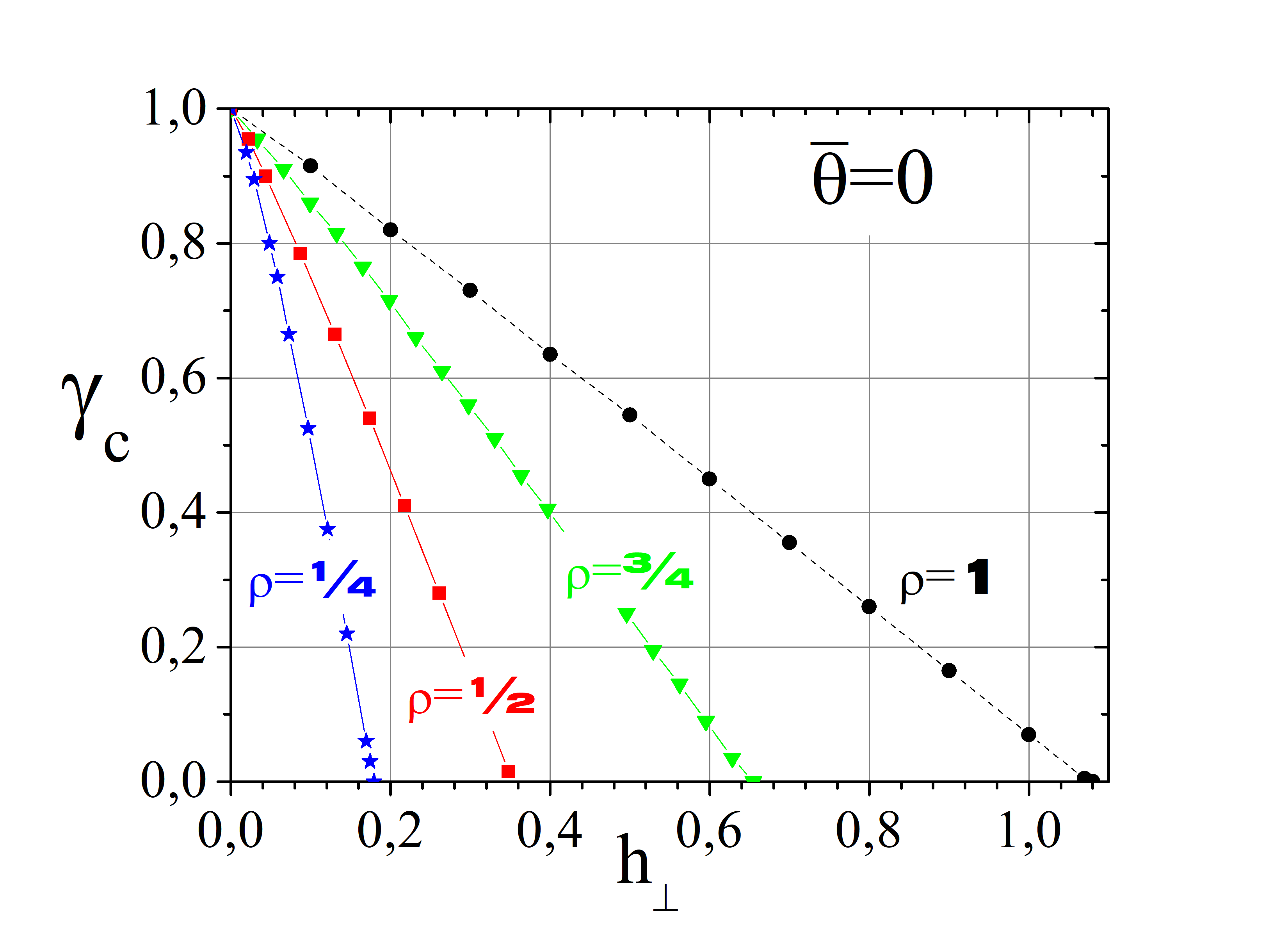}}
\subfigure[ ]{\includegraphics[width=8cm]{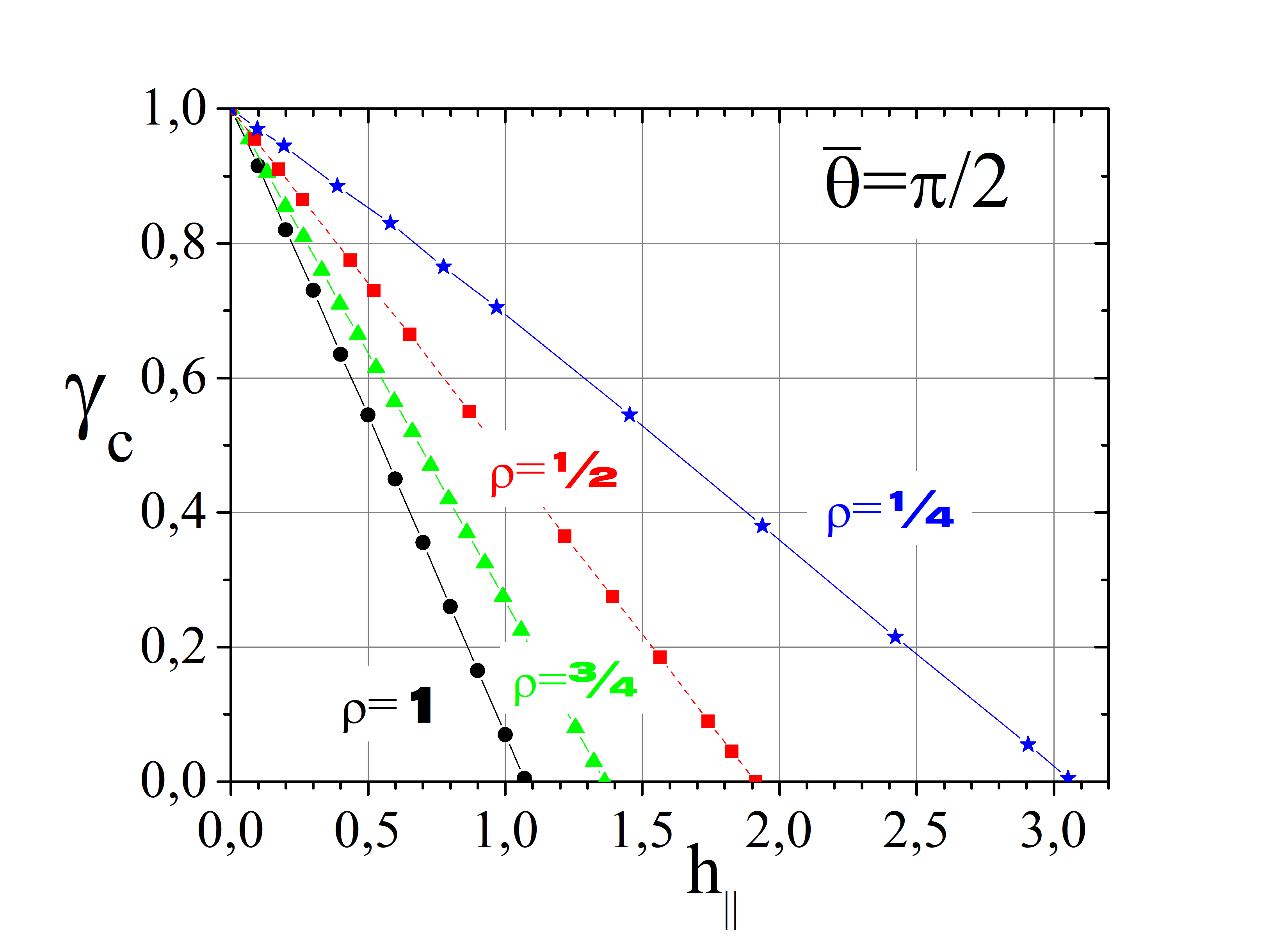}}
\caption{(Color online) Numerically computed magnetic diffraction patterns, $\gamma_c(h)$, of a one-dimensional CAJTJ with $\ell=50$ and $n=0$ for different values of the aspect ratio, $\rho$, and two values of the in-plane field orientation, $\bar{\theta}$, relative to the annulus major diameter: (a) $h_{\bot}$ for $\bar{\theta}=0$, and (b) $h_{\parallel}$ for $\bar{\theta}=\pi/2$. The magnetic field strength is normalized to $J_c/ c \cosh\bar{\nu}$.}
\label{numMDPs}
\end{figure}
 
\subsection{Static simulations}

\noindent To begin with, numerical integrations of Eq.(\ref{psge}) have been carried out in the stationary, i.e., time-independent, state ($\phi_{\hat{t}}=0$) to derive the magnetic diffraction pattern (MDP) of the critical current of the CAJTJs. Specifically, we have numerically computed the maximum (or critical) value, $\gamma_c=I_c(H)/I_c(0)$, of the normalized zero-voltage current versus the normalized field amplitude, $h=H/J_c c\cosh\bar{\nu}$ with the initial phase profile $\phi_{\tau}=0$ for the normalized bias current $\gamma_0=0$ in Eq.(\ref{psge}); then $\gamma_0$ was ramped-up in small increments of $0.01$ and the phase profile recorded until a stationary, i.e., time-independent solution exists. Strictly speaking a uniform initial phase profile only allows for the determination of the first or main lobe of the $\gamma_c(h)$ pattern. We considered two orthogonal orientations of the in-plane magnetic field relative to the annulus major diameter: a field $h_{\bot}$ perpendicular to the major axis corresponds to a field orientation $\bar{\theta}=0$ in the magnetic forcing term $F_h$ defined in Eq.(\ref{Fh}). Vice-versa, for $\bar{\theta}=\pi/2$ the field is parallel to the major diameter and will be named $h_{\parallel}$. The numerically computed field dependencies, $\gamma_c(h_{\bot})$ and $\gamma_c(h_{\parallel})$ are shown in Figs.~\ref{numMDPs}(a) and (b), respectively. As the MDPs are symmetric, $\gamma_c(-h)=\gamma_c(h)$, we only consider positive fields. We observe that all the plots are approximately linear but have quite different slopes, i.e., different critical fields $h_c$, the values at which the main lobe of the MDP first goes to zero, so that $\gamma_c(h_{c})=0$. We remind that, with our field normalization, the critical current of an infinitely long circular ($\rho=1$) AJTJ corresponds \cite{JAP07} to $h_c=1$. Since our simulated rings have a large, but finite, normalized perimeter, $\ell=50$, the critical	field is slightly larger than unity, $h_c(\rho\!=\!1)\approx1.08$; as a circle has infinitely many axes of symmetry, this occurs for any field orientations. From the figures it is seen that, as the annulus is made more and more eccentric, the critical field decreases (increases) when the in-plane applied field is perpendicular (parallel) to the major axis. For the most squeezed confocal annulus, $\rho=1/4$, the ratio of the parallel to perpendicular critical field is about $18$, i.e., much larger than the ratio of the major to the minor axis of the outer ellipse which is very close to $3$ (see Fig.~\ref{ConfAnn}). The reason of such markedly different effects resides in the fact that the $2\pi$-periodic (polarity-dependent) magnetic potential breaks the symmetry of the width-induced double-well potential unless when the field is strictly perpendicular to the major diameter of the CAJTJ.

\subsection{Dynamical simulations}

\begin{figure}[t]
\centering
\subfigure[ ]{\includegraphics[width=8cm]{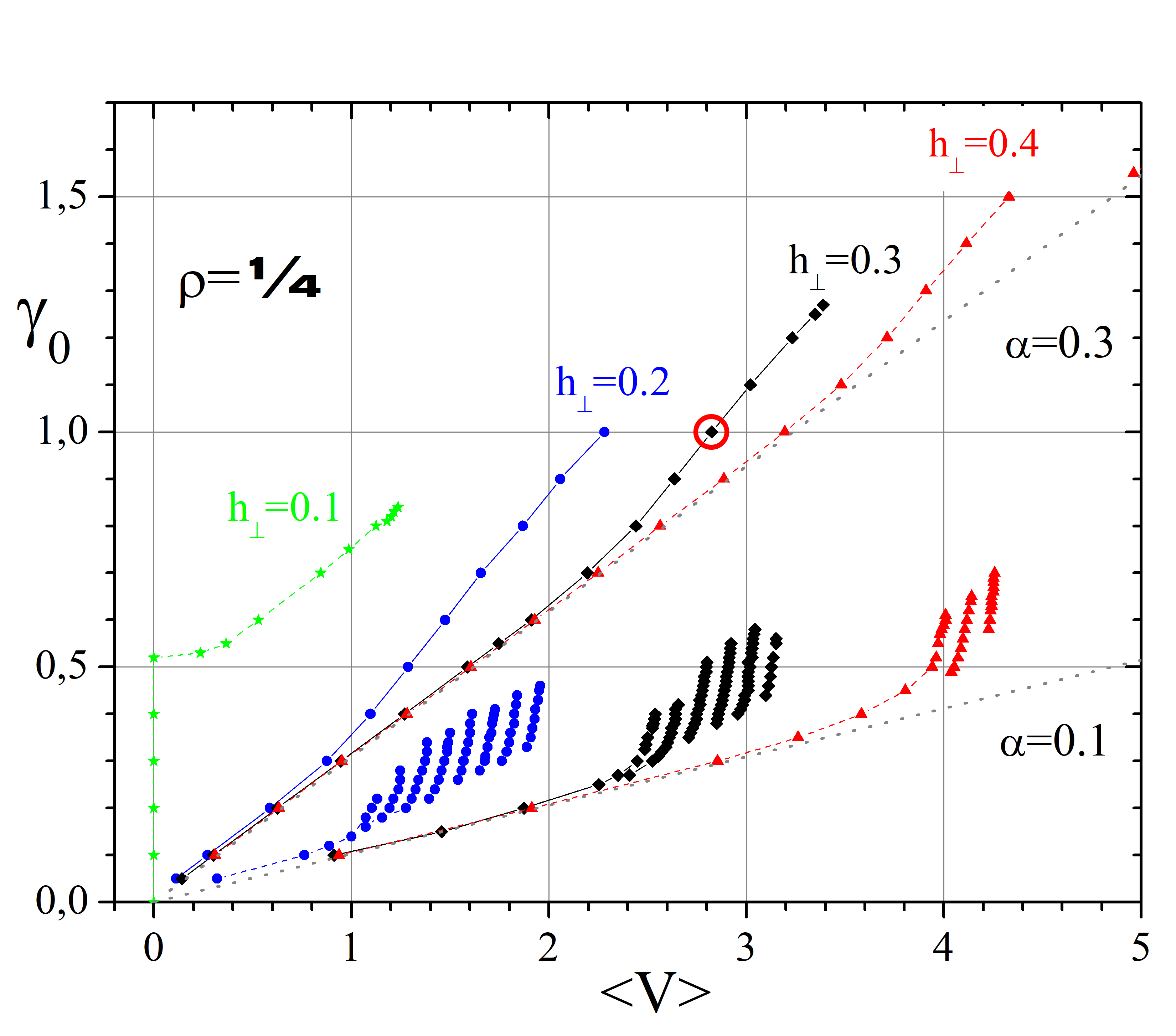}}
\subfigure[ ]{\includegraphics[width=8cm]{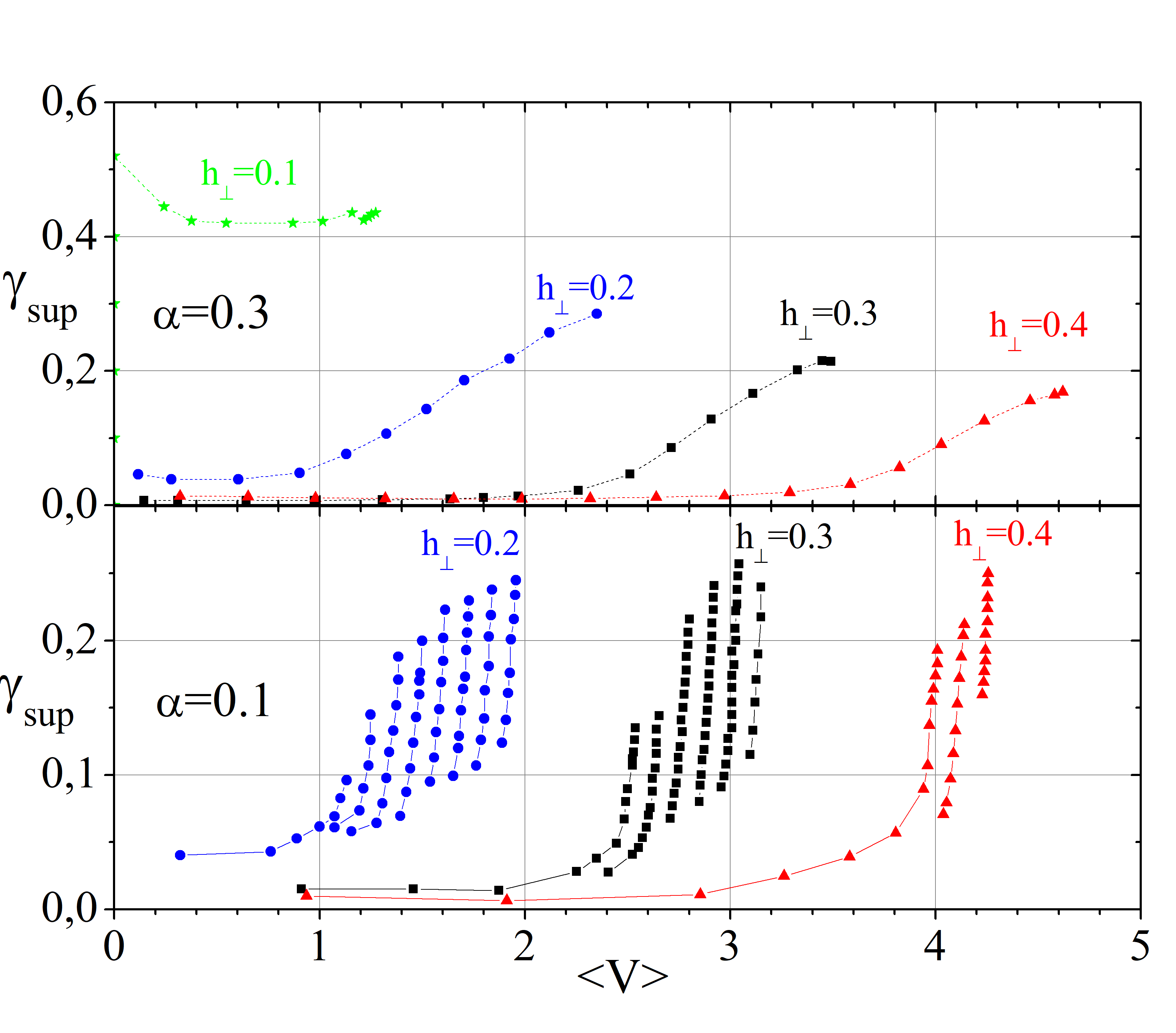}}
\caption{(Color online) (a) Numerically computed current-voltage characteristics of a CAJTJ with aspect ratio $1\!:\!4$ and normalized length $\ell=50$ obtained by fixing the loss parameter $\alpha$ and varying the value of perpendicular magnetic field $h_{\bot}$, as indicated by the labels. The dotted lines indicate the ohmic current, $\gamma_{nor}=\alpha <\!\!V\!\!>$. (b) as in (a) but with the background ohmic current, $\gamma_{nor}\equiv\alpha <\!\!V\!\!>$, subtracted; the bottom panel refers to $\alpha=0.1$ and the top panel to $\alpha=0.3$.}
\label{IVCnum}
\end{figure}

\begin{figure}[b]
\centering
\subfigure[ ]{\includegraphics[height=7cm,width=8cm]{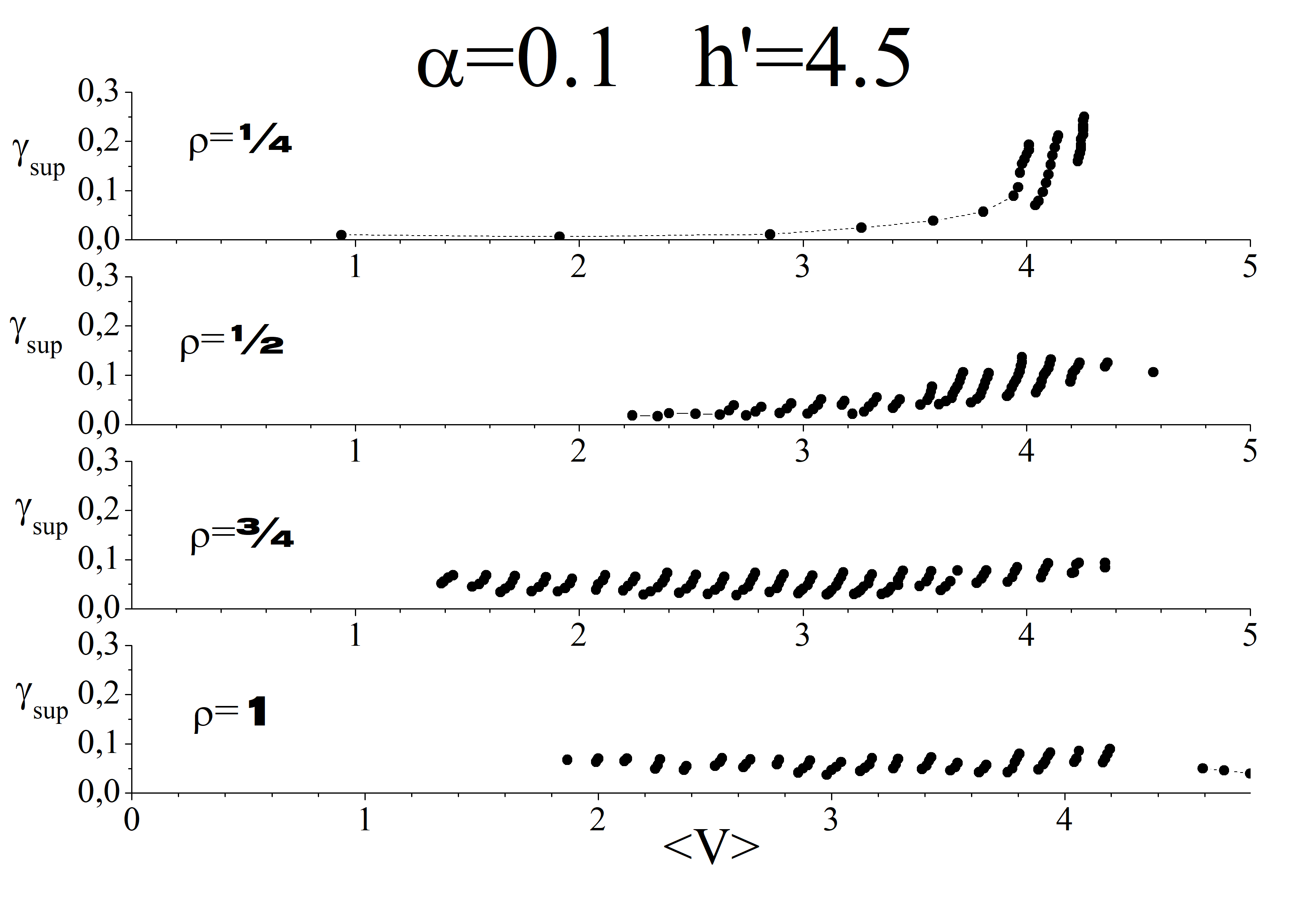}}
\subfigure[ ]{\includegraphics[height=7cm,width=8cm]{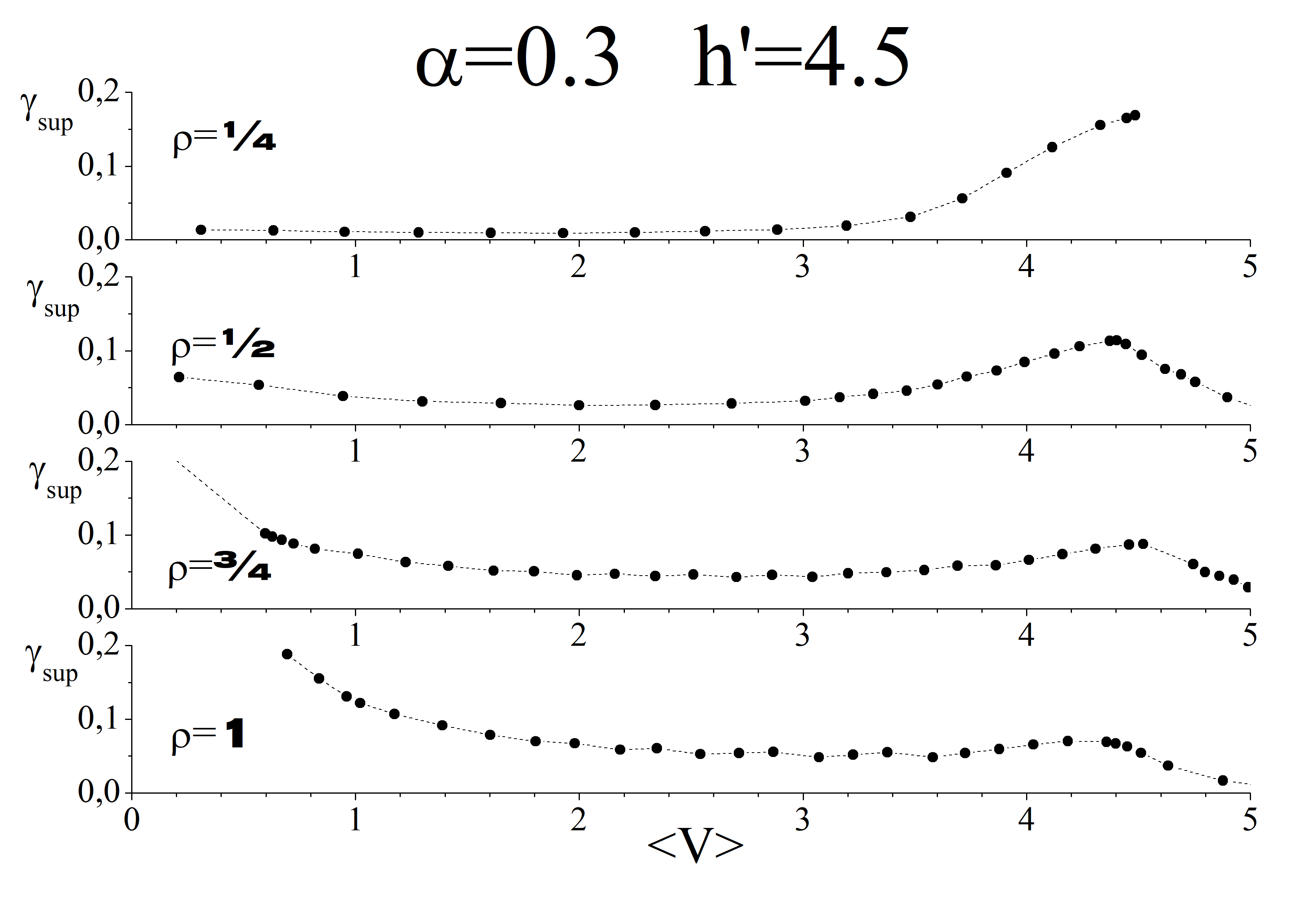}}
\caption{Numerically computed current-voltage characteristics of a CAJTJ with normalized length $\ell=50$ obtained by fixing the value of the perpendicular magnetic field $h^\prime_{\bot}\equiv H_{\bot}/J_c \lambda_J=4.5$ and varying the aspect ratio as indicated by the labels. The simulations were carried out for two values of the loss parameter $\alpha$: (a) $\alpha=0.1$ and (b) $\alpha=0.3$ (see text).}
\label{FFS}
\end{figure}

\noindent Fig.~\ref{IVCnum}(a) shows the numerically computed current-voltage characteristics of a CAJTJ with aspect ratio $1\!:\!4$ and normalized length $\ell=50$ obtained for two values of the loss parameter, $\alpha=0.1$ and $0.3$, and three values of the perpendicular magnetic field strength, $h_{\bot}=0.2$, $0.3$ and $0.4$. The dotted lines indicate the ohmic current, $\gamma_{nor}=\alpha <\!\!V\!\!>$. Each point in the plots corresponds to a flux-flow dynamical state whose time evolution will be considered later on. Such solutions are periodic in time and space and their frequency, $2\pi/T$, with $T$ being the time periodicity, is identified with the average voltage, $<\!\!V\!\!>$, that could also be evaluated by averaging $\phi_{\hat{t}}(\tau,\hat{t})$ over a sufficiently long time. It is seen that the IVCs markedly depend on the loss parameter, $\alpha$. For $\alpha=0.3$, a DLS is observed at $h_{\bot}=0.1$ which is below the critical value $h_{\bot,c}\approx0.26$. At field strength, $0.2$, slightly below the critical value, a smooth and continuous ES is found whose voltage, for a given current, depends linearly with the value of the external magnetic field. Note that, for $\alpha=0.3$, the product $\alpha \ell=15$ is well above $2\pi$. The situation seems to change when $\alpha \ell$ is close to or smaller than $2\pi$. Indeed, for $\alpha=0.1$, the numerically computed FFSs consists of a set of steep and equally spaced high-order FSs; their voltage separation $\Delta\!<\!\!V\!\!>$ is about $0.12$, i.e., close to $2\pi/\ell=0.128$ that is the asymptotic voltage of the first ZFS calculated when one fluxon is trapped in the AJTJ ($n_w=1$). The width of each single Fiske resonance is approximately equal to $\alpha$ which explains why for $\alpha$ larger than $\Delta\!<\!\!V\!\!>$ we enter the parameter space region where it is not possible to distinguish resonances anymore and we observe only a smooth and continuous singularity. This effect is better observed in the two panels of Fig.~\ref{FFS}(b) where the same data of Fig.~\ref{FFS}(a) have been replotted in terms of the supercurrent, $\gamma_{sup}\equiv\gamma_0-\gamma_{nor}$ that is computed as the spatio-temporal average of $\sin \phi(\tau,\hat{t})$ and provides information on the stability of the dynamical state. Our numerical investigation indicated that, regardless of the loss parameter, the voltage position of the numerically computed steps increases with the field approximately as $h^\prime_{\bot}\equiv h_{\bot} c \cosh \bar{\nu}/\lambda_J=H_{\bot}/J_c \lambda_J$; incidentally, $h^\prime \equiv H/J_c \lambda_J$ is the magnetic field normalization typical of linear long linear JTJs \cite{Cirillo98}, whose critical field is $h^\prime=2$. The amplitudes of the step show a weak field dependence, however, for a given field, the step heights drastically reduces as the junction eccentricity is lowered. This is shown in Figs.~\ref{FFS}(a) and (b) that report the IVCs computed for different aspect ratios, $\rho$, with $\alpha$ equal to, respectively, $0.1$ and $0.3$. In order to have the same maximum voltage of the branches, the field strength was set to $h^\prime_{\bot}\approx 4.5$ that corresponds to the value $h_{\bot}=0.4$ used in Figs.~\ref{IVCnum} for $\ell=50$ and $\rho=1/4$. In both cases we found that the FFS tend to disappear as the eccentric annuli change into rings. This is consistent with our initial observation that the JFF has never been reported for the well-studied circular AJTJs. In passing, we note that, in Fig.~\ref{FFS}(b), due to the relatively high value of the loss parameter, the resonant nature of the Eck step is clearly seen, since it is possible to trace the negative resistance part of the curve.

\noindent So far we have presented numerical results in the presence of a perpendicular field simply because the resonances excited by a parallel field are infinitesimally small; more specifically, as the direction, $\bar{\theta}$, of the applied field is rotated from $0$ to $\pm\pi$, the supercurrent of the magnetically-induced branches continuously decreases until they almost disappear in the ohmic background currents. This is consistent with the absence of magnetically induced structures noticed in the experiments.

\subsection{The Josephson flux-flow state in AJTJs}

\begin{figure}[b]
\centering
\includegraphics[width=8cm]{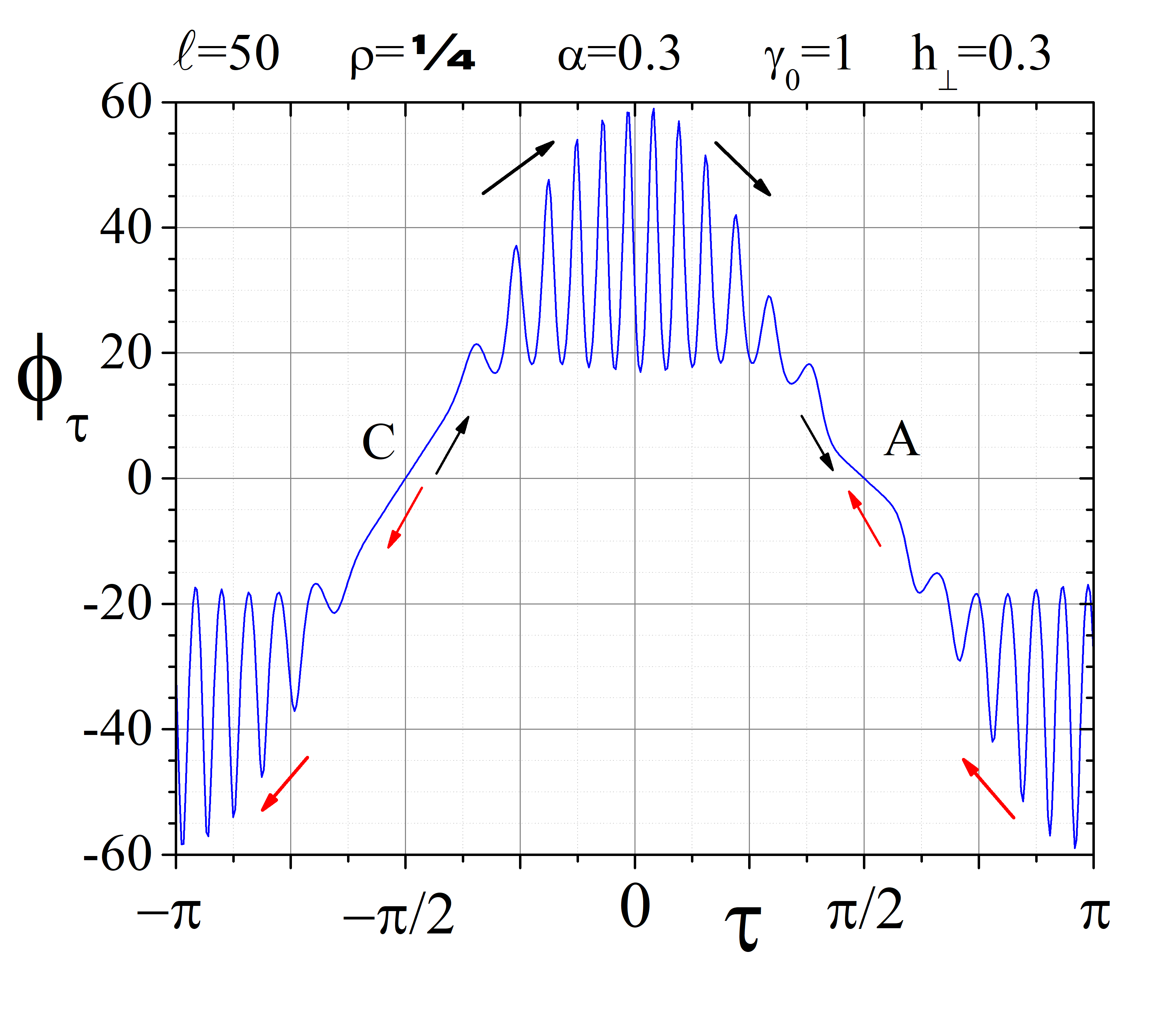}
\caption{(Color online) $\phi_{\tau}$-profile obtained for $\rho=1/4$, $\alpha=0.3$, $\gamma_0=1$ and $h_{\bot}=0.3$ that corresponds to the point marked by an open circle in Fig.~\ref{IVCnum}(a) with $\gamma_{sup}\approx0.13 $.}
\label{ux}
\end{figure}

\noindent Generally speaking, when an in-plane magnetic field with arbitrary orientation is applied to an unbiased long AJTJ, some magnetic flux penetrates the tunnel barrier at the two diametrically opposite points where the tangential field component is largest (in absolute value). Depending on the field direction and strength, a number of static $F\bar{F}$ pairs are accumulated on one annulus side, while, for symmetry reasons, the same number of static $\bar{F}F$ pairs are stored in the diametrically opposite side. In the presence of a bias current applied to the AJTJ, both the fluxons and antifluxons experience a Lorentz force the direction of which depends on their polarity. As a result, depending on the current sign, on one side the Lorentz force pushes the static fluxons and antifluxons against each other and annihilate them, while on the other side, more interestingly, the fluxons and the antifluxons start to propagate along the annulus perimeter until they eventually collide after traveling half a turn. The result of the collision between a fluxon and an antifluxon moving at a given velocity in opposite directions drastically depends on the loss of the system. Indeed, on a lossless line, the kinks survive the collision without change of shape, speed or trajectory regardless of their kinetic energy. In the presence of dissipative effects, a threshold velocity exists \cite{Scott78} above which the kinks pass through each other without mutual destruction. Below the threshold, the kinks fade off by the breather decay mode in which the fluxon-antifluxon pair is bound together in a damped oscillatory state \cite{Nakajima74}. At last, the kinks with the opposite polarity annihilate each other and their energy is partly radiated onto the line and partly dissipated by some loss factors contained in the line. The threshold velocity increases with the losses. It was demonstrated that in long JTJs the surface-impedance loss severely reduces the threshold velocity \cite{Ruang91}, especially when the shunt loss and bias current are small \cite{Pedersen84}.

\noindent The analysis of the time evolution of the numerically computed solutions of Eq.(\ref{psge}) enabled us to understand the mechanism underlying the JFF in AJTJ and the conditions which enhance or weaken the process. Indeed, all the magnetic resonances, DLSs, FSs, ESs, reported in this section rely on just one common flux-flow steady-state dynamics that is qualitatively illustrated by means of Fig.~\ref{ux} that shows the profile of the spatial phase derivative, $\phi_{\tau}$, taken at an arbitrary time. In the presence of a perpendicular field, $F\bar{F}$ pairs are continuously created at the left equatorial point, $\tau=-\pi/2$, pinpointed by the letter $C$, with a rate proportional to the field strength. Under the influence of the Lorentz forces due to the bias current and the magnetic field, the fluxons (positive pulses) rotate clockwise (increasing $\tau$), as indicated by the black arrows, while the antifluxons (negative pulses) rotate anticlockwise (decreasing $\tau$), as indicated by the red arrows. Since, for symmetry reasons, they travel with opposite but equal speed, they collide at the diametrically opposite equatorial point, $\tau=-\pi/2$, identified by the letter $A$. If the fluxons and antifluxon created in $C$ collide in $A$ at sufficiently small velocity, they annihilate at the same rate at which they were initiated and a robust flux-flow state is developed with well separated kinks and a large supercurrent. This is the case of the $\phi_{\tau}$-profile in Fig.~\ref{ux} obtained for $\rho=1/4$, $\alpha=0.3$, $\gamma_0=1$ and $h_{\bot}=0.3$ that corresponds to the point marked by an open circle in Fig.~\ref{IVCnum}(a) with $\gamma_{sup}\approx0.13 $. It is seen that, in this specific case, about 12 fluxons and 12 antifluxons are involved in the JFF state; as they move in a complex spatial potential, they have a position-dependent speed, and different widths and amplitudes above the almost sinusoidal background. The data show that an average phase difference of $\Delta \phi\approx\pm 78.5$ exists between the creation and annihilation points that yields a more accurate evaluation of the average number, $\Delta \phi/2\pi\approx12.5$, of kinks participating in the flux-flow in each semi-annulus. Their velocity is smallest near the creation and annihilation points and largest around the poles ($\tau=0$ or $\pm \pi$). The average voltage of this dynamical state, $<\!\!V\!\!>\approx2.8$, divided by the number of kinks involved in the process, provides an estimation of the average fluxon speed when compared to the asymptotic voltage associated with just one traveling fluxon, namely, $0.112/0.128\approx87\%$ of the Swihart velocity. Increasing the bias current, both the average number and the average speed of the kinks increase. A complete annihilation is the necessary requirement for a stable flux-flow process. The bias current that supplies energy to the fluxons and the losses which subtract energy certainly play a determinant role and a balance must be achieved. However, above all, the eccentricity of the CAJTJ is crucial: in fact, the fluxon-antifluxon annihilation is strongly favored when it occurs in the well of the width-dependent fluxon potential of very eccentric confocal AJTJs, that is exactly what happens in the presence of a perpendicular in-plan field. As the confocal annulus tends to a ring, the potential well disappears (see Fig.~\ref{Qmat}) and the annihilation becomes less likely. The JFF process is less attainable (if not impossible) in the presence of a parallel magnetic field, in which case the fluxons-antifluxons collision occurs in one of the metastable points at $\tau=0$ or $\pm \pi$, where the fluxon potential has a relative maximum.

\section{Comments and conclusions}

The comparison between the experimentally recorded and the numerically computed families of IVCs reveal a more than satisfactory qualitative agreement. A number of reasons may explain the quantitative discrepancies. First, the adopted model does not include the effect of the bias-current induced self-field which is particularly strong in high-$J_c$ samples with a current flow perpendicular to the direction of the applied magnetic field. In addition, the uniform bias approximation is not realistic for our ``in-line like'' geometrical configuration for which a current distribution, $\gamma(\tau)$, peaked at the equatorial points, $\tau=\pm \pi$, would be more appropriate. Above all, the voltage-independent loss parameter, $\alpha$, is responsible for the luck in the simulated families of IVCs of a seamless transition from the Fiske-staircase to the smooth ES as the steps move away from the current axis with the increasing magnetic field strength. Despite these caveats, however, our study clearly elucidate the conditions under which the viscous flow of Josephson vortices can occur in long AJTJs. The flux-flow process manifest itself through finite-voltage structures, such as DLS, FSs and ESs, in the current-voltage characteristics induced by an in-plane magnetic field. The width of such resonances is determined by the inverse of the system ohmic loss $\alpha$. Unlike the case of flux flow in a type-II superconductor \cite{Kim65}, where a critical magnetic field must be exceeded, such a critical condition has not been observed for the flux flow in AJTJs. In fact, several effects attributed to vortex motion, such as the DLSs, have been observed in an external magnetic field smaller than the first critical field of the supercurrent; this holds even more true in the experiments where some of the magnetic field is provided by a dc bias current of several milliamperes. The vortex dynamics for a given magnetic field has revealed that a fluxon train  with internal degrees of freedom travels in one half of the annulus perimeter while a train of antifluxons moves in the opposite half. A robust steady flux-flow involves the fluxon-antifluxon annihilation inside the junction and requires that the system is not very underdamped. However, also geometrical parameters were found to drastically affect the flux-flow in AJTJ. The well-studied circular annular configuration could not support a consistent JFF. On the contrary, the confocal AJTJs which are the natural generalization of the circular AJTJs, allows for very stable flux-flow states due to the intrinsic non-uniformity of their planar tunnel barrier delimited by two closely spaced \conf ellipses. The richer nonlinear phenomenology of CAJTJs provides an elegant example of how the geometrical subtleties are of paramount importance in the physics of \Jos tunnel junctions. More specifically, magnetically induced structures carrying a large supercurrent, which measure the robustness of the flux-flow state, have been observed in CAJTJs with large aspect ratio ($\rho=1/2$ or smaller) provided that the in-plane uniform magnetic field is applied perpendicular to the junction major axis. Under this conditions the motion in opposite directions of the fluxon and antifluxon trains is symmetric and the annihilation occurs in one of the wells of the fluxon double-well potential intrinsic to the periodically changing junction width. Our experimental findings with samples of quite different geometrical and electrical parameters as well as the numerical simulations made over a large parameter space indicate that any deviation from the above conditions worsen the quality of the JFF. 

  


%

%


%


\section*{Acknowledgments}
\noindent RM acknowledges the support from the Italian CNR under the Short Term Mobility Program 2018. RM and JM acknowledge the support from the Danish Council for Strategic Research under the program EXMAD. The fabrication of the experimental samples was carried out at IREE within the framework of the state task and partially supported by the Russian Foundation for Basic Research, grant No 17-52-12051.

\newpage

\end{document}